\documentclass[pra,floatfix,preprint,nofootinbib,eqdisecnum,12pt]{revtex4}
\usepackage{epsfig}
\usepackage{bm}
\usepackage{amsmath}
\usepackage{setspace}
\input epsf
\numberwithin{equation}{section}

\def\ed{ed.~by~}

\def\HP{$(H,\rho_F,\rho_I)$}

\def\H{H}

\def\rhoI{\rho_I}
\def\rhoF{\rho_F}
\def\cN{{\cal N}}

\def\be{\begin{equation}}
\def\ee{\end{equation}}

\begin{document}
\vspace{1cm}

\title{Arrows of  Time and  \\ Initial and Final Conditions in the \\  Quantum Mechanics  of Closed Systems\\ Like the Universe}
\author{James B.~Hartle}

\email{hartle@physics.ucsb.edu}

\affiliation{Department of Physics, University of California,Santa Barbara, CA 93106-9530}
\affiliation{Santa Fe Institute, Santa Fe, NM 87501}

\date{\today }

\begin{abstract} 
\singlespacing
A model quantum cosmology is used to illustrate how arrows of time emerge in a universe governed by a time-neutral dynamical theory $(H)$  constrained by time asymmetric initial and final boundary conditions represented by density matrices $\rhoI$ and $\rhoF$.  In a quantum universe universe arrows of time are described by the probabilities of appropriately  coarse grained sets of histories of quantities like entropy that grow or decay.  We show that the requirement of that these sets of histories  decohere implies two things: (1) A time asymmetry between initial and final conditions that is a basis for arrows ot time.  (2) How a final state of indifference that is represented by a density matrix $\rhoF$ proportional to the unit density matrix is  consistent with causality, and allows a finer-grained description of the model universe in terms of decoherent histories than any other final state.
\end{abstract}





\maketitle

\bibliographystyle{unsrt}


\section{Introduction}
\label{intro}

Contemporary theories governing  the dynamics of the universe are usually assumed to be time-neutral --- not distinguishing one time direction over any other. A familiar example is a CPT invariant quantum field theory coupled to a classical cosmological geometry.  Yet our universe exhibits various time asymmetries defining arrows of time. Familiar examples are:

\begin{itemize}

\item The thermodynamic arrow of time --- the fact that
approximately isolated systems are now almost all evolving towards equilibrium
in the same direction of time.


\item The arrow of time of retarded electromagnetic
radiation.


\item The arrow of time defined by  the
expansion of the universe.

 \item The arrow of time supplied by the growth of fluctuations away from initial   inhomogeneity and isotropy with the universe's expansion . 
 
 \end{itemize}
 More examples of arrows of time are discussed in \cite{96}. 
 
 The first three of the  arrows of time above 
can be in principle be reversed temporarily, locally, in isolated subsystems,
although typically at an expense so great that the experiment can
be carried out only in our imaginations.  If we could, in the
familiar example of Loschmidt \cite{Los1877}, 
reverse the momenta of all particles and
fields of an isolated subsystem, it would ``run backwards'' with the
thermodynamic  arrow reversed. We cannot of course time reverse our universe.
 
 The disparity between the time symmetry of the fundamental laws of
physics and the time asymmetries of the observed universe has been a
subject of fascination for physicists since  the late 19th century and the literature on the subject is vast.  For a sample of this literature, including a number of reviews see, e.g. \cite{Los1877,Dav76,Pen78,Gri84,Joos85,Zeh89,HLL93,HH2-12,Har2-13,Har2x19}.

 Both in quantum mechanics and classical statistical physics  these time asymmetries could arise from time-symmetric dynamical
laws constrained by  time-asymmetric boundary conditions.  When there is a well defined notion of time it is conventional to call one of these boundary conditions the `initial condition'  and the other a `final condition'.   A thermodynamic
arrow of time, for example, would be  implied by an initial condition in
which the progenitors of today's approximately isolated systems were all far from
equilibrium at an initial time and a final condition of indifference at a later final time. 
We could say that the thermodynamic arrow of time {\it emerged} \cite{Har19} from a time symmetrical dynamical framework due to time asymmetric boundary conditions\footnote{Note that we are not claiming that all arrows of time exhibited by the universe arise by this mechanism. See, e.g. \cite{BKM14}.}.

The time evolution of a quantum system is not generally described by a single history of how events happen in time as it might be in classical physics..
Rather it is described by a set of alternative possible histories with quantum probabilities for which occurs.  The inputs to calculating these probabilities are first, a dynamical theory which we denote by $(H)$ and assume time neutral. Initial and final boundary conditions  represented  by density matrices $\rhoI$ at a time $t_I$  usually assumed pure, and a density matrix $\rhoF$ at time $t_F$ which is the subject of this paper.

The universe displays an arrow of time when the probabilities are high for histories  that describe the systematic growth (or decay) of a  physically interesting quantity defined at a series of  times. A suitably coarse grained entropy, or the amount of retarded electromagnetic radiation are examples. We are therefore interested in formulation of quantum mechanics that does not just predict  probabilities of alternatives at particular moments of time but rather a quantum mechanics of time histories.  We will use the consistent or decoherent histories formulation of quantum theory. (DH). References to its foundations can be found in \cite{classicDH}.  A bare boned description of the parts essential for this paper can be found in Sections \ref{model} and \ref{time-neutral-qm}. Throughout we assume that the universe is spatially closed. 

DH predicts probabilities that are consistent with the rules of probability theory only for sets of alternative histories for with there is negligible quantum interference between the individual histories in the set. Such a set of histories is said to decohere.  

This paper is concerned with the limitations on $\rhoF$ and $\rhoI$ arising from the requirement that the set of alternative histories describing arrows of time decohere and therefore the limitations on their time asymmetry.
We have two findings:

(1) There is no decoherent set of histories in which the $\rhoF$ is the same as a pure initial state  $\rhoI$.  Decoherence requires time asymmetry in the boundary conditions for prediction in closed quantum systems. This is less surprising given that typical mechanisms of environmental decoherence assume disapation and therefore  an arrow of time   \cite{CL83,GH93a}.  One should therefore not be surprised that our  universe exhibits arrows ot time.

(2) The requirement of decoherence means that a final condition of indifference $\rhoF\propto I$  allows a finer-grained description of the universe in terms of decoherent histories than any other final state.  The reason can be simply stated. Decoherence requires coarse graining --- following some variables describing the universe but ignoring others.  Interaction of the followed variables with  the  ``environment'' of the ignored variables produces the decoherence of sets of histories in the followed variables as in many discussions of environmental decoherence show e.g \cite{JZ85,Zur03,GH13,GH93}.The interaction creates {\it records} are that are strongly correlated by with the  individual histories in the environment  but orthogonal to each other. That orthogonality produces decoherence. It is a remarkable fact that in DH quantum theory it is necessary to lose some information in order to have any information at all. 

(3) In quantum mechanics arrows of time do not generally arise just from  special initial conditions alone as has sometimes been suggested.  Rather the arrows arise  by differences between initial and final conditions. Some arguments emphasizing the role of special initial conditions can be interpreted as implicitly assuming a particular initial condition $\rhoI$ and  a final condition of indifference $\rhoF   \propto I$, where $\rhoF$ is proportional to the unit matrix $I$ so that arrows arise from the differences between initial and final conditions 

  In quantum cosmology  it is usual  to assume a pure state for the initial condition so that $\rhoI=|\Psi\rangle\langle \Psi|$. The no-boundary wave function of the universe \cite{HH83, HHH18} is an well studied  proposal for $|\Psi\rangle$. For the final condition it is common to assume  a condition of indifference  meaning that  $\rhoF$ is proportional to the unit density matrix $\rhoF\propto I$. In simple models this combination has successfully predicted  arrows of time along with many other features of our large scale universe. e.g \cite{HLL93,HH2-12,Har2-13}.
 
 
The purpose of this paper is not to present a complete explanation of the  emergence of the universe's time asymmetries. Rather it is to exhibit a prerequisite for such discussions for  a model quantum universe. Specifically we discuss constraints on the final quantum conditions that arise  from decoherence in a time neutral  decoherent histories formulation of a quantum mechanics of  a closed quantum system like we have assumed for our universe.  We  then  use such constraints to address the question of why we should assume a final condition of indifference. 

 To keep the discussion manageable we will restrict attention to a simple model  of a closed system.  This is a large,  cosmologically sized, box,  
 perhaps expanding, and containing particles and fields as suggested in  Figure $1$  and specified in more detail in 
 Section \ref{model}. Everything is contained within the box --- galaxies, planets, the Earth, Sun, and Moon, observers and
observed, measured subsystems, and any apparatus that measures them, you and me. There is nothing outside and no influence of the outside on the inside or the inside on the outside.  In this simple model of a quantum universe,  arrows of time, if any, are properties of the probabilities  predicted by such a quantum theory for the individual members of decoherent sets of alternative coarse-grained time histories of what goes on {\it inside}  the box. For example, if the probability is generically high for histories in the set in which  fluctuations grow from an initial to final time a  fluctuation arrow of time is predicted.

 To discuss the emergence of time asymmetries from differences between initial and final conditions $\rhoI$ and               $\rhoF$ we need not only a time neutral dynamical theory  but also a time neutral formulation of quantum mechanics with no built in arrows of time. That way the arrows of time will be emergent from the theory and not posited in the formulation of quantum theory. This paper is mainly concerned with that framework and its consequences for prediction.

 The familiar textbook (Copenhagen) formulation quantum mechanics for measurements cannot  serve this purpose.  It  has a built in time asymmetry.  Unitary evolution by the Schr\"odinger equation can be run both forward and backward in time.  But the reduction of the state on measurement works in only one direction in time  thus specifying a built in arrow of time.. This is often assumed to be in the direction of the classical thermodynamic arrow of time.  (The author knows of no compelling justification for this assumption, and the experimental evidence for it is limited at best.). We need a generalization of Copenhagen quantum mechanics that is time neutral and free from any such built in arrows of time. 
 

An appropriate time-neutral quantum framework for  closed systems  is already available in the time-neutral decoherent (or consistent) histories formulation of quantum theory \cite{96}  incorporating insights  from  e.g.  \cite{JZ85,ABL64,Gri84}, and the principles of generalized quantum theory e.g. (\cite{Har19}, Section 4).  The  ingredients are,  first  a Hamiltonian $H$ specifying quantum dynamics.   
Second, there  are both initial  and final conditions specified by  density matrices
$\rho_I$ and $\rho_F$ at times $t_I$  and $t_F$ respectively. (To simplify the notation we will often suppress the times  $t_I$ and $t_F$ 
writing just $\rho_F$ for $(\rho_F, t_F) $ for example. 
The theoretical input to the calculation of probabilities of alternative histories of the closed system is therefore the triple $(H,\rhoI, \rho_F)$.
The conditions $\rhoI$ and $\rhoF,$  enter this formalism symmetrically so there are no built in quantum arrows of time. Differences between $\rhoI$ and $\rhoF$ can lead to physical arrows of time. The thermodynamic arrow is an example. When the entropy of $\rhoI$ is low at $t_I$  and the entropy of $\rhoF$ is high at $t_F$,  as it is for example  when $\rhoF$ is proportional to the unit density matrix $I$, we can expect a thermodynamic arrow of time to emerge. 

A pure initial state $\rhoI = |\Psi\rangle\langle \Psi|$ is a natural candidate for the initial condition because, if he universe is a quantum system, it has some quantum state. The no-boundary quantum state \cite{HH12,HH83,HHH18}  is a natural, well explored, candidate for the initial state were we dealing with cosmology including quantum gravity.   A condition of indifference $\rho_F \propto I$ where $I$ is the unit density matrix  is oft assumed as a final condition. Such a final condition is simple, generally consistent with  causality,  and familiar from the text book (Copenhagen) quantum mechanics  of measurement situations.  But what is argument for this final condition  in cosmology?   

This paper uses the simple model to show that the requirement of decoherence means that a final condition of indifference allows a finer-grained description of the universe in terms of decoherent histories than any other final state.  The reason can be simply stated. Decoherence requires coarse graining --- following some variables describing the universe but ignoring others.  Interaction of the followed variables with  the  ``environment'' of the ignored variables produces the decoherence of sets of histories in the followed variables as in many discussions of environmental decoherence show e.g \cite{JZ85,Zur03,GH13,GH93}.The interaction creates {\it records} are that are strongly correlated by with the  individual histories in the environment  but orthogonal to each other. That orthogonality produces decoherence. It is a remarkable fact that in DH quantum theory it is necessary to lose some information in order to have any information at all. 

A final state of indifference $\rhoF\propto I$  has zero information, it doesn't specify anything in particular. It could be said to be as coarse-grained as possible leaving as many degrees of freedom as possible to enable  decoherence. By contrast, as we will show explicitly  that a pure final state
$\rhoF=|\chi\rangle\langle\chi|$ prohibits any decoherence necessary for quantum probabilities for histories.

The structure of the rest of the paper is as follows: As an  easily understood example Section \ref{EMarrow} shows how the familiar retardation of electromagnetic arises from an initial condition of negligible  free radiation in the early universe that would evolve to  detectable radiation today an a final condition of indifference to how much free radiation there is in the far future. Section \ref{model} introduces our model universe in a box. Section \ref{time-neutral-qm} describes the essential parts of DH that will be needed in the subsequent argument in particular the quantitative measure of decoherence. Section \ref{noteclass} sketches the modifications of the argument that would be needed were spacetime geometry treated quantum mechanically (quantum gravity). Section \ref{conclusion} is a brief conclusion.

\section{The Classical  Arrow of Time of Retarded Electromagnetic Radiation}
\label{EMarrow}
The arrow of time defined by the retardation of classical electromagnetic radiation  provides a simple example of an arrow  arising  
 from time-asymmetric cosmological boundary conditions applied to
time-reversible dynamical laws\footnote{Parts of this section are adapted from \cite{Har05b}.}. The time reversible dynamical laws  are Maxwell's equations for 
the electromagnetic field in the presence of charged sources in the universe. 
 
The time-reversal invariance of Maxwell's equations implies that {\it
any} solution for specified sources  can be written at
in {\it either} of two ways. First,  (R) a sum of a free field (no
sources) coming from the past plus {\it retarded} fields whose sources are
charges in the past. The second is  (A) a sum of a free field coming from the future
plus {\it advanced} fields whose sources are charges in the future.
More quantitatively, the four-vector potential
$A_\mu(x)$ at a point $x$ in spacetime can be expressed in the presence of
four-current sources $j_\mu(x)$ in Lorentz gauge as either
\begin{eqnarray*}
A_\mu(x)&=A^{\rm in\ }_\mu (x) + \int d^4 x^\prime D_{\rm ret}
(x-x^\prime)\, j_\mu (x^\prime)\quad &(R) \\
\noalign{\hbox{or}}
A_\mu(x)&=A^{\rm out}_\mu (x) + \int d^4 x^\prime D_{\rm adv}
(x-x^\prime)\, j_\mu (x^\prime) \quad &(A).
\end{eqnarray*}
Here, $D_{\rm ret}$ and $D_{\rm adv}$ are the retarded and advanced Green's
functions for the wave equation and  $A^{\rm in}_\mu (x)$ and $A^{\rm out}_\mu
(x)$ are free fields defined by these decompositions. When the sources $j_\mu(x)$ are limited to a bounded
range of time, $A_\mu^{in}(x)$ describes source-free electromagnetic radiation in the distant past.  Similarly 
$A_\mu^{\rm out}(x)$ describes source-free radiation in the far future. 

Suppose there were no free electromagnetic fields in the distant past 
so that $A^{\rm in\ }_\mu(x)\approx 0$.
Using the R description above, this time asymmetric boundary condition
would imply that present fields can be entirely ascribed to sources in the
past. This is retardation and that is the electromagnetic arrow of time.

The advanced free field $A^{\rm out}_\mu(x)$ is determined from relation  (A) above  at late times once  $A_\mu(x)$ is known from (R) and  $A_\mu^{in}(x)$ is known or predicted..

Thus, we could say that the electromagnetic arrow of time emerges from the 
special initial condition of  $A^{\rm in\ }_\mu (x) \approx 0$ and a final condition of indifference as
to what $A^{\rm out}_\mu(x)$ turns out to be.



The expansion of the universe has red-shifted the peak luminosity of the 
CMB at decoupling to microwave wavelengths today. There is thus a 
negligible amount energy left over from the big bang in the wavelengths 
we use for vision, for instance.  A contemporary human observer functioning
at wavelengths where the CMB is negligable  will therefore be receiving 
information about charges in the past.  This selection of 
wavelengths is plausibly not accidental but adaptive \cite{Har05b}. A   
contemporary observer seeking to function with input from microwave
wavelengths would find little emission of interest, and what there
was would be overwhelmed by the all-pervasive CMB, nearly equally bright in
all directions, and carrying no useful information.



\begin{figure}[t]
\label{boxfig}
\includegraphics[width=5in]{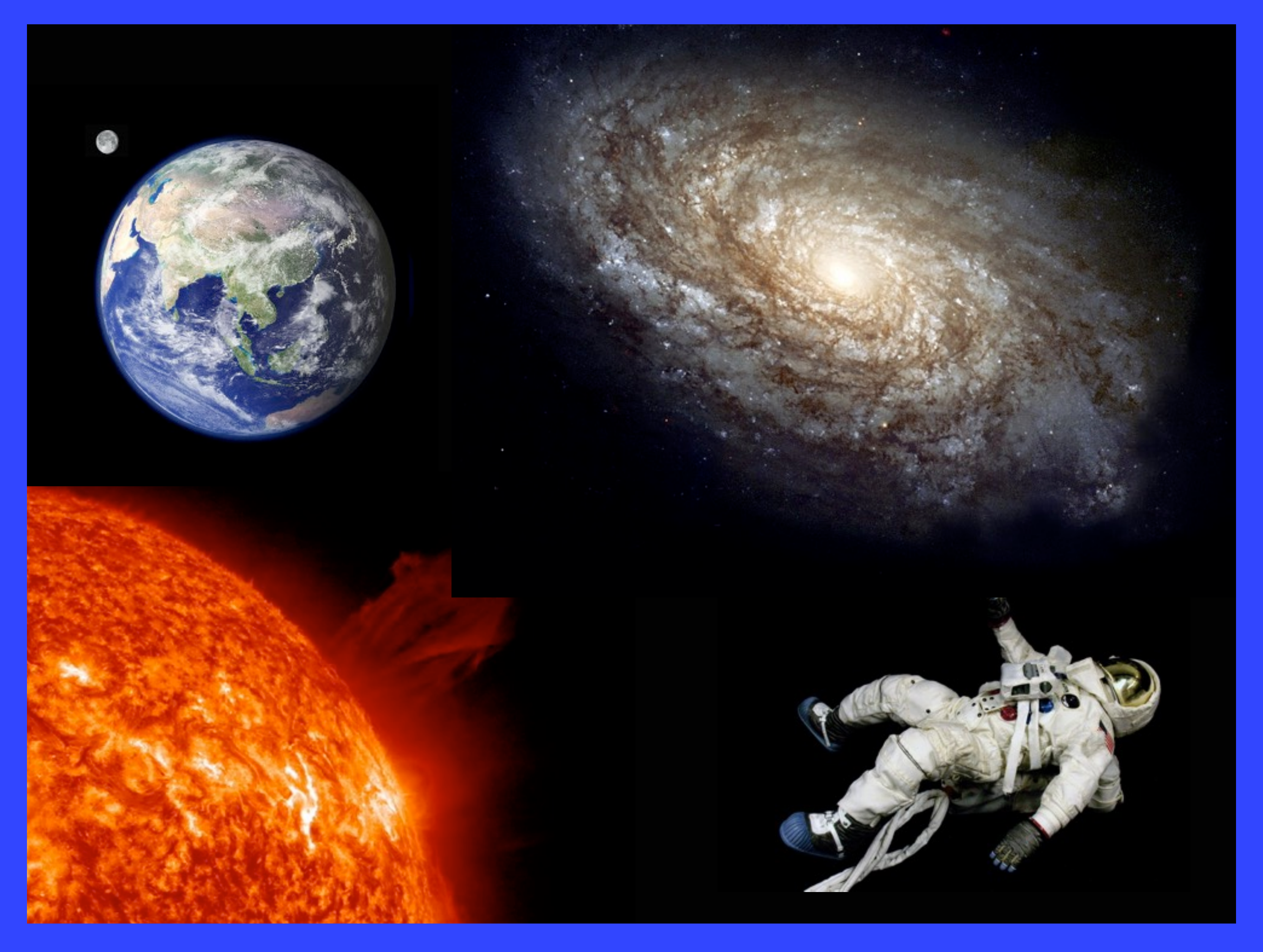}\hfill
\caption{A simple model of a closed quantum system is a universe of quantum matter fields inside a large closed box (say, 20,000 Mpc on a side) with fixed flat spacetime inside.  Everything is inside the box --- galaxies, stars, planets, human beings, observers and observed, subsystems that are  measured and subsystems that are measuring. The most general objectives for prediction are the probabilities of the individual members of decoherent sets of alternative coarse grained histories that describe what goes on in the box. That includes histories describing any measurements that take place there. There is no observation or other meddling with the inside from outside.  }
\end{figure}
\section{A Model Quantum Universe in a Closed Box}
\label{model}

 We  assume  a fixed, flat, background spacetime inside our model  box thus neglecting quantum gravity. This an excellent approximation in the realistic universe for times later than a very short interval  $\sim 10^{-43}sec.$ after the big bang. As a consequence there  is  a well defined notion of time in any particular Lorentz frame. The familiar apparatus of textbook quantum mechanics then applies ---  a Hilbert space, operators, states,  and their unitary evolution in time.  We assume a quantum field theory in the flat spacetime for dynamics. Everything is in the box and there is no interaction between its inside and outside.

The basic theoretical inputs for predicting what goes on in the box    the  Hamiltonian $\H$ governing time evolution and the initial and final density matrices   $\rhoI$  and $\rhoF$ written here in the Heisenberg picture for convenience. Input theory is  then \HP. Oftenwe assume that the initial density matrix is pure $\rhoI=|\Psi\rangle\langle.$  The basic output are probabilities for what goes on in the box. More precisely the outputs of \HP\ are probabilities for the individual members of sets of coarse-grained alternative time histories of what goes on in the box. The examples in the next sections  will make the important ideas introduced in this sentence  more concrete.

\subsection{A very large experiment} 
\label{expt}

 It may help some readers to imagine that our box is part of a very large experiment constructed  by observers with far more resources than we have. The observers {\it prepare}  the box in an initial pure quantum state $|\Psi\rangle$ at a time $t_I$.  At a later time $t_F$ they {\it select} an ensemble final states weighted by the probabilities in a density matrix $\rhoF$.  

Such a  final selection will necessarily influence what histories are predicted between $t_I$ and $t_F$. That is acausal  --- the action taken in the future that influences what occurs in the present and past. Or we could say that by an analysis of what doesn't occur in the present we could predict what does have to happen in the future.  For example, in certain circumstances we could find information about the final condition from present observations.  The weak measurements discussed by Aharonov, Vaidman, and others \cite{twostate} are other examples.  Not unrelated  examples are in  cosmology are \cite{Cr95} and \cite{Har94}. 

Causality could be assumed were  $\rhoF \propto I$ because then there is no final selection at all.

\subsection{Coarse Grained Histories of the Universe}
\label{cgcosmo}

In cosmology we are interested in the probabilities of decoherent sets of   coarse-grained alternative classical histories that describe the universe's expansion, the primordial nucleosynthesis of light elements, the formation and  evolution of the microwave background radiation, the formation  of the galaxies, stars, planets, the evolution of biota,  etc. Such sets of alternative histories relevant for our observations in are highly coarse grained. They don't  describe everything that goes on in the universe --- every galaxy, star, planet, human history, etc., etc in all detail. Rather they follow  much coarser grained histories of the universe. 

In laboratory science we may be  interested in histories that describe the preparation, progress, and outcomes of a particular measurement situation.  In our box model quantum cosmology there are no measurements of  the inside of the box by something outside it.  Laboratory measurements are described realistically,  as a correlation between one subsystem inside the box that includes the apparatus, observers, etc and another subsystem inside that is thus measured. In this way measurements can be described in the quantum mechanics of the universe but play no
preferred role in the  formulation of its quantum mechanics as they do in Copenhagen quantum theory. Probabilities for the outcomes of measurements are particular instances of the probabilities that describe what goes on in the universe e.g. \cite{Har91a}. 

 \subsection{Coarse-grained Histories}
\label{cghist} 
To understand the relevant histories we focus on a simple example:  the set of histories which 
 describe the positions of the Moon in its motion around the Earth at a series of times $t_1,\cdots, t_n$.  We are then interested in the probabilities of the   alternative orbits that the  could Moon follow around the Earth.  Each orbit is an example of a {\it history} --- a sequence of events at a series of times. 
Relevant histories are {\it coarse-grained}  because we are only interested in positions defined to an accuracy consistent with our observations of the Moon's center of mass position, and further  because positions are not specified at each and every time but only at a finite discrete sequence of times. Coarse grained histories can be said to {\it follow} certain variables and {\it ignore} others. In the present example the histories follow the center of mass of the moon and ignore variables that describe the interior of the Moon and Earth. In quantum mechanics there is no certainty that the coarse-grained history of the  Moon's center of mass  will follow a classical Keplerian orbit, but in the Moon's situation the probability predicted by \HP\ is vastly higher for a classical Keplerian orbit than for a non-classical one. 

\section{A Time Neutral Decoherent Histories Quantum Mechanics of the Universe }
\label{time-neutral-qm}

 This section  presents a bare bones description of how the theory \HP\  predicts probabilities for which of a decoherent  set of alternative coarse-grained histories happens in the model box. Many more details and specific models can be found in e.g. \cite{classicDH,Har95c}. 
 
 \subsection{Histories} 
 \label{histories}
 
 For simplicity let's  continue to focus on the motion of the Moon. 
 The simplest  set of histories  describing the motion of the Moon is obtained  by giving a sequence of  yes-no alternatives at a series of times. 
For example: Is the center of mass of the Moon in a region $R$ of the box at this time --- yes or no?  This alternative is {\it coarse-grained}  because it does not  follow Moon's position exactly but only whether it is in $R$ or not. The alternative `yes' is represented by the projection operator $P_R$ on the region $R$ amd `no' by $I-P_R$.  More generally coarse-grained yes-no  alternatives at one moment of time are described by an exhaustive set of exclusive Heisenberg picture,  projection operators $\{P_\alpha(t)\}$, $\alpha=1,2,3,\cdots$ acting on the Hilbert space of the Moon'a center of mass.   These satisfy:
\be
\label{projections1} 
P_\alpha(t)P_{\alpha'}(t) = \delta_{\alpha\alpha'} P_\alpha(t), \quad \sum_\alpha P_\alpha(t) = I . 
\ee
showing that the projections are exclusive and exhaustive. 

Projections on bigger subspaces are more coarse grained, projections on smaller subspaces are finer grained. 

In the Heisenberg picture in which we work, projection operators representing the same alternative  at different times are connected by unitary evolution defined by the Hamiltonian $H$, viz.  
\be
P_\alpha(t') = e^{iH(t'-t)/\hbar}P_\alpha(t) e^{-iH(t'-t)/\hbar} . 
\label{un-evol}
\ee

For example, to  describe the quasiclassical realm of every day experience inside the box the relevant projections would be products of projections  for each subvolume onto ranges of values of the quasiclassical variables  --- averages  over suitable volumes of energy, momentum, and number e.g. \cite{GH07}. 

A set of alternative coarse-grained histories of the Moon'a center of mass  is specified by a sequence of such sets of orthogonal projection operators at a series of times $t_1,t_2, \cdots t_n$. An individual history of the Moon's orbit corresponds to a particular sequence of events  $\alpha \equiv (\alpha_1,\alpha_2, \cdots, \alpha_n)$ and is represented by the corresponding chain of projections:
\be
C_\alpha\equiv C_{\alpha_n\alpha_{n-1} \cdots \alpha_1}  \equiv P^n_{\alpha_n}(t_n) \dots P^1_{\alpha_1}(t_1) .
\label{class}
\ee
where $\alpha$ is a is a shorthand for the chain of $\alpha's$  in  \eqref{class}.
As $\alpha_1,\alpha_2, \cdots, \alpha_n$ range over all possible values a {\it set} of coarse-gained alternative  histories of the Moon's center of mass  is defined. We denote the set of alternative histories  by 
$\{C_\alpha\}$. Quantum theory aims at predicting the probabilities  that these alternative histories occur.

\subsection{Decoherence and Probabilities}
\label{decohandprob}

A wide range of  generalizations of Copenhagen quantum theory can be constructed by specifying two things:  First, the set of coarse-grained histories ${\{C_\alpha\}}$ of interest and second a measure of the quantum interference between any pair of coarse-grained histories in the set ${\{C_\alpha\}}$. 

The measure of quantum interference between any pair of histories $(\alpha', \alpha)$ in a set $\{C_\alpha\}$  is called the {\it decoherence functional}  and is denoted by $D(\alpha',\alpha)$. 
This must  defined to incorporate appropriate notions of positivity, hermiticity, nnormalization,  and be with the principle of superposition  \cite{Har91a}.  A set of alternative coarse-grained histories {\it decoheres} when the off diagonal elements of $D$ are negligible. The diagonal elements are then the predicted probabilities. $p(\alpha)$  for the histories, viz.
\be\boxed{
\label{decoherence}
D(\alpha',\alpha) \approx  \delta_{\alpha'\alpha} \ p(\alpha).}
\ee

Eq.\eqref{decoherence} is the central equation in any decoherent histories formulation of quantum theory. It specifies both when a set of coarse grained alternative histories $\{C_\alpha\}$ decoheres, and what the probabilities $p(\alpha)$ of the individual members of the set of histories are. These probabilities are the predictions of the theory. Decoherence is a necessary condition for the $p(\alpha)$ defined by \eqref{decoherence} to be {\it consistent}  with the usual rules of probability theory \cite{Gri84}.

To complete the specification of the quantum framework it remains to specify the decoherence functional.  
For the time neutral formulation of quantum mechanics with initial and final conditions we take \cite{96}
\be\boxed{
\label{decohfnl}
D(\alpha',\alpha) \equiv \cN Tr[C^{\dagger}_{\alpha'}\rhoF C_\alpha \rhoI], \quad  1/\cN \equiv Tr(\rhoI \rhoF).}
\ee
This decoherence functional  is time neutral in the sense that using the cyclic property of the trace $\rhoI$ and $\rhoF$ can be interchanged.
Thus they enter the formalism symmetrically so that there is no built in arrow of time as there is in the textbook quantum mechanics of measurement situations.

Specializing to a pure initial state $\rhoI \equiv |\Psi\rangle\langle\Psi|$ this  becomes
\be
\label{decohpure}
D(\alpha',\alpha) / {\cN} \equiv \langle \Psi  | C^{\dagger}_{\alpha'}\rhoF C_{\alpha}|\Psi\rangle= \langle\Psi_{\alpha'} |\rhoF |\Psi_{\alpha}\rangle
\ee
where,  for convenience,  we have defined  branch state vectors corresponding to the individual coarse-grained histories 
\be
\label{branches}
|\Psi_\alpha\rangle\equiv C_\alpha |\Psi\rangle .
\ee
With the assumption of a pure $\rhoI$  the decoherence condition becomes (supressing time labels)
\be
\label{geom}
\boxed{\langle\Psi_{\alpha'} |\rhoF |\Psi_{\alpha}\rangle =\delta_{\alpha'\alpha}.}
\ee
This has a simple geometrical interpretation in David Craig's geometry of consistency \cite{Cr97}.
We can think of $\rhoF$ as a positive metric on the space of histories.  A set of histories decoheres if their branch state vectors are mutually orthogonal in the metric supplied by the final condition $\rho_F$. 

\section{Final Condition Limitations  on the Number of Decoherent Histories.}
\label{limits-final} 

To begin a discussion limitations on histories arising from final conditions we consider two limiting cases. First a final condition of indifference $\rho_F \propto I$ and then the opposite condition of a pure state final condition.  $\rho_F=|\chi\rangle\langle\chi|$ for some $|\chi\rangle$.
\subsection{A Final Condition of Indifference}
\label{indif}
When $\rho_F\propto I$ the decoherence condition \eqref{geom} becomes $\rangle\chi|\langle\chi|$
\be
\label{decoh}
D(\alpha',\alpha) / \cN \equiv  \langle \Psi |  C^{\dagger}_{\alpha'} C_{\alpha}|\Psi\rangle= \langle\Psi_{\alpha'} |\Psi_{\alpha}\rangle\approx \delta_{\alpha',\alpha} =\langle\Psi_{\alpha'} | \Psi_\alpha\rangle .
\ee
The set of histories decoheres if the branch state vectors are mutually orthogonal.  If the dimension of the Hilbert spce is $N$ we can have a maximum of $N$ histories in a decohering set. A final condition $\rhoF \propto I$ allows the maximum number of decohering histories consistent with the dimension of the Hilbert space.  If the Hilbert space is infinite dimensional there is no restriction at all. 

\subsection{No Non-Trivial Decoherence with a Pure State Final Condition}
\label{pure}

 Suppose the initial condition is pure a pure state  $\rhoI=|\Psi\rangle\langle\Psi |$ and 
 suppose that  the final condition is also  a pure state  $\rho_F\equiv|\chi\rangle\langle\chi | $ for some state $|\chi\rangle$ so that the rank of the density matrix is unity. Then, 
\be
\label{pureF}
D(\alpha',\alpha)/\cN =\langle\chi | C_{\alpha'}|\Psi\rangle^*\langle\chi |C_\alpha|\Psi\rangle.
\ee

Number the histories $\alpha=1, 2, \cdots$. To have a decohering set of histories at least one of 
$\langle\chi | C_{\alpha'}|\Psi \rangle$ must be non-zero. Suppose that is $\alpha'=1$. For the set of histories to decohere all the $\langle\chi | C_{\alpha'}|\Psi\rangle$  must vanish for $\alpha'>1$. Thus, a pure final state allows only sets with one history (and its negation) to decohere. Assuming a pure final state results in only a single history being predicted with unit probability which is no prediction at all. 

\subsection{The General Case}
\label{general}
The general case should now be clear.  Suppose $\rhoF$  has $K<N$ eigenvectors with non-zero eigenvalues.  Then the theory \HP\  can predict probabilities for $K$ different histories. This result is consistent with the special cases above. For a final condition of indifference $K=N$ and for a pure state final condition $K=1$.  

\subsection{Decoherence Requires Time Asymmetry}
\label{time-asym}
The discussion in the above sections shows that if the initial density matrix  is pure $\rhoI\equiv |\Psi\rangle\langle\Psi|$ then the final state cannot be also pure and support non-trivial sets of decohering histories. In particular the final density matrix $\rhoF$ cannot be the same as the initial one $\rhoI$.  Time asymmetry between initial and final conditions is thus necessary for decoherence. The physical reason was discussed in the introduction. Information has to be lost to effect decoherence in all but trivial sets of  alternative histories\footnote{Even if $\rhoI$ is not pure there are still stringent conditions that must be satisfied for a set histories with the same initial and final density matrix. See Section VIB in \cite{96}.}.

Therefore we should not be surprised that our universe exhibits arrows of time. The prerequisite time asymmetry is a natural consequence  of  the formulation of quantum theory of closed systems and the mechanisms of decoherence. 



\subsection{Causality}
\label{causality} 
\label{experiment} 

Think for a moment about an even larger experiment  with many boxes  of the kind  in Section \ref{model}. The observers {\it prepare}  each in an initial quantum state $|\Psi\rangle$at at  time $t_I$.  At a later time $t_F$ they {\it select} final states weighted by the probabilities in a density matrix $\rhoF$.

Then the final condition in the future would influence what occurs in the present and past.  This is acasusal.\footnote{The weak measurements discussed by Aharonov, Vaidman, and others \cite{twostate} are examples. Not unrelated  examples in  cosmology are \cite{Cr95} and \cite{Har94}. }.

To build in causality to the basic theory restrict   $\rhoF \propto I$ because then there is no final selection at all.

\section{ A Note on Quantum and Classical Spacetime}
\label{noteclass}
The model universe in a box introduced in Section \ref{model}, and used throughout, assumed a (flat) classical spacetime  inside the box. A classical spacetime is a central assumption in textbook (Copenhagen) quantum mechanics.  It defines the time in the Schr\"odinger equation as well as the family of spacelike surfaces through which quantum states evolve unitarily. Classical spacetime is also effectively assumed in the time neutral, decoherent histories formulation of quantum theory discussed in Section \ref{time-neutral-qm} with which we analyzed initial and final conditions at definite moments of time $t_I$ and $t_F$.

The evidence of the observations is that  in our universe contains a suitably coarse-grained classical spacetime  which extends over the whole of the visible universe from a very short time after the big bang to a little before the big crunch singularity in a recollapsing universe or to the indefinite future when there is no such singularity. In a realistic theory of quantum cosmology, the singularity theorems of classical cosmology show that there is no classical spacetime near  the big bang. And if the classical universe recollapes to a future singularity (the big crunch) there won't be classical spacetime immediately before\footnote{At a big crunch one could investigate whether there could be  a quantum transition to a further regime of classical spacetime.}.

The classical behavior of anything is not a given in a quantum universe. It is  a matter of quantum probabilities.  A quantum system behaves classically when, in a suitably coarse-grained set of alternative histories, the probabilities are high for  for histories exhibiting correlations in time governed by deterministic classical laws, for example  by the Einstein equation (e.g. \cite{GH07,GH93,HHH08}). A generalization of quantum theory  that does not assume classical spacetime is  thus necessary to discuss the emergence of classical spacetime in the early universe. We need a formulation of quantum mechanics which can supply probabilities for histories of spacetime geometry to predict when and where is a domain of classical spacetime. Frameworks for such generalizations have been sketched  which do address the emergence of classical spacetime e.g \cite{Har95c,Hal11}.  The no-boundary quantum state has been applied in the semiclassical approximation to predict domains of classical spacetime  and other realistic features of our universe such as the amount of inflation e.g. \cite{HHH08}. The quantum formalisms used  have the analogs of initial and final conditions but to the author's knowledge the kind of analysis represented in this paper remains to be carried out  for these cases.

\section{Conclusion}
\label{conclusion}
\subsection{The Main Points Again}
\label{mainpoints}
\begin{itemize}
\item{The universe exhibits an arrow of time when the quantum probability is high that  the history of some physically interesting quantity like a suitably defined entropy or the amount of retarded electromagnetic radiation increases or decreases in time generally and systematically.}

\item{In a quantum universe the growth or decay of any quantity is described by the probabilities of sets of alternative suitably coarse-grained histories that track the possible evolution of the relevant quantities over time. }

\item{The probabilities of the histories in the relevant set depend on a theory of quantum dynamics $H$ and theories of the  initial and final conditions $\rhoI$ and $\rhoF$ respectively.}. 

\item{In a quantum theory arrows of time  arise from time asymmetries between initial and final conditions and not just  just special initial conditions as sometimes assumed. }  

\item{For the predicted probabilities of the  histories in a set of alternative ones to be consistent with the rules of probability theory there must be negligible quantum interference between individual histories in the set. That is, the set of histories must decohere.}

\item{When $\rhoI$ is a pure state the requirement of decoherence means that $\rhoF$ cannot be pure and must be different from $\rhoI$ creating a time asymmetry in boundary conditions that is possible  origin of arrows of time.  We should therefore not be surprised that our universe exhibits arrows of time.}

\item{For given $\rhoI$ and $H$ a final condition of indifference $\rhoF \propto I$ leads to finer grained description of the universe through  a larger number of decoherent histories than with any other $\rhoF$.} 

\item{The theory $(H,\rhoF,\rhoI)$ is tested by the probabilities it predicts for features of the universe that we observe, among these features are the various arrows of time the it predicts.}

\end{itemize} 

\subsection{An Unfinished Task of Unification?}
\label{unfinished}
In the context of our model box universe with a pure initial state we have shown that a final condition of indifference $(\rhoF\propto I)$ has two theoretically attractive features. First, it is consistent with an elementary notion of causality.
Second, it allows a finer grained description of the universe in terms of decoherent sets of alternative histories than other final conditions.  

However, it goes without saying, that  the initial and final conditions and  the theory of dynamics and the quantum framework in which they are applied are not decided by  such theoretical argument. They are discovered by of comparing the predictions of different theories  $(H,\rho_I, t_I, \rho_F, t_F$)   with large scale observations of our universe.  

To the author this state of affairs suggests that there is an unfinished task of unification. The dynamics, and the initial and final conditions of our model box universe are independently specifiable. We hope that for our unique quantum universe there is a unique, unified set of principles that determines all of these?\footnote{Already the no-boundary wave function of the universe \cite{HH83,HHH18} can be seen as unifying the dynamics and the initial quantum state.}. 

\acknowledgments  
This  paper was written in part to combat the ennui due to a stay in hospitals arising from a hip fracture and during the ensuing recovery period. The author thanks the staff who cared for him and above all  his wife Mary Jo for her care, advice,  and companionship through this trying time. 
 The author acknowledges many discussions with Murray Gell-Mann$^\dagger$, Thomas Hertog, and Mark Srednicki  on the quantum mechanics of the universe. He thanks the Santa Fe Institute for supporting many productive visits there. This work was supported in part by the US. National Science Foundation under grants PHY15-04541 and PHY18-18018105 .


\end{document}